\documentclass[aps,pra,floatfix,amsmath,ams,superscriptaddress,showpacs,12pt]{
revtex4-1}
\usepackage{amsmath,amssymb}
\usepackage{amsfonts}
\usepackage{mathrsfs}
\usepackage{epsfig}
\usepackage{graphicx}
\usepackage{dcolumn}

\def\be{\begin{equation}}
\def\ee{\end{equation}}

\begin{document}
\title{Entanglement transitions in random definite particle states}

\author{Vikram S. Vijayaraghavan}
\email{vikram@ms.physics.ucdavis.edu}
\altaffiliation{Current address: Department of Physics, University of California, 
One Shields Avenue,
Davis, CA 95616-8677.}
\author{Udaysinh T. Bhosale}
\email{bhosale@physics.iitm.ac.in}
\author{Arul Lakshminarayan}
\email{arul@physics.iitm.ac.in}
\affiliation{Department of Physics, Indian Institute of Technology Madras, Chennai, 600036, India}
\date{\today}
\preprint{IITM/PH/TH/2010/11 }
\begin{abstract}
Entanglement within qubits are studied for the subspace of definite
particle states or definite number of up spins. A transition from
an algebraic decay of entanglement within two qubits with the total number $N$ of qubits, to an exponential one when the number of particles is increased from two to three is studied in detail. In particular the probability that the concurrence is non-zero is calculated using statistical methods and shown to agree with numerical simulations. Further entanglement within a block of $m$ qubits is studied using the log-negativity measure which indicates that a transition from algebraic to exponential decay occurs when the number of particles exceeds $m$. Several algebraic exponents for the decay of the log-negativity are analytically calculated. The transition is shown to be possibly connected with the changes
in the density of states of the reduced density matrix, which has a divergence at the zero eigenvalue
when the entanglement decays algebraically. 

\end{abstract}

\maketitle

\section{Introduction}

Entanglement has been extensively investigated in the recent past, as
it is a critical resource for quantum information processing
~\cite{nielsenbook}. One model of quantum computation, the one-way
quantum computer \cite{Raussendorf01}, relies explicitly on entanglement.  The resource of
entanglement is not at all rare, a random pure quantum state is
typically highly entangled \cite{bandyo03,Hayden06}. In fact there is
so much entanglement in typical random pure states that recent studies
\cite{bremner09,Gross09} find them not to be useful for one-way
quantum computation. This motivates the question of studying subsets
of states with a control over the amount of entanglement available.

It is well known that most of the entanglement in many body quantum
systems is multipartite. In random pure states of $N$ qubits, we need
to consider blocks whose total size is at least larger than $N/2$, for
them to be entangled \cite{kendon02a}. This being the case,
entanglement in smaller blocks is nearly impossible to observe.
Previous studies have shown how rare it is to have two qubits
entangled in a many qubit random pure state \cite{scott03,kendon02a}.
In this paper it is shown that there is a surprising connection
between the number of up-spins or particles present in definite
particle states and entanglement. Thus producing definite particle
random states, as defined below, may allow control over the type of
entanglement that is desired. For instance if two qubit entanglement
is to be obtained, it is shown that typical three-particle states will
render this nearly impossible to achieve. The border between probable
and improbable is described by a transition from an algebraic to an
exponential decay, which is typically obtained at phase
transitions. Further, the approach presented in this paper might
  shed light on methods that are applicable to a wider class of
  problems in the area of quantum complex systems.

Random pure states, or ``full" random pure states, belong to the ensemble of states that are uniformly
sampled from the Hilbert space, with the only constraint being
normalization, in other words sampled from the unique Haar measure.
Such states arise for instance in mesoscopic systems
\cite{beenakker97}, nuclear physics~\cite{brody81} etc.  and have been
modeled as eigenfunctions of random matrices from the usual Gaussian
ensembles \cite{MehtaRMT}. There have been studies that explore how to efficiently
produce operators with statistical properties of random matrices
\cite{emerson03}. Classically chaotic systems have long been known to
exhibit such states in their quantum limit, and studies of
entanglement in quantum chaotic systems often take recourse to random
states~\cite{benenti08,arul01}.

The ensemble of interest in this work is taken to be the one where
{\it all} the vectors that are constrained to be in the definite particle subspace are equally
likely, subject again only to the constraint of normalization. This is equivalent to assuming that
the Hamiltonian in the definite particle symmetry reduced subspaces are full random matrices.
In this paper the random matrices are taken to be of the GOE (Gaussian Orthogonal Ensemble) \cite{MehtaRMT}
type and hence the states studied are real. Any study that 
uses random states or random matrices is justified as a baseline with which to compare
realistic Hamiltonian systems that may involve interactions in a complex,
nonintegrable manner. In particular it is possible that due to the few-body nature of typical 
interactions, ensembles such as the Embedded GOE \cite{Kota01} will be of interest. 
However we believe that studying a usual ensemble like the GOE 
will form a baseline for entanglement studies of  a rather 
large class of physically important systems which conserve particle number,
or total spin. 
 
A definite particle state is a random pure state in a fixed $S_z$ subspace formed
by the basis vectors of the Hilbert space, which, when expressed in
the spin-$z$ basis, have a fixed number, say $l$, of ``ones", or spin
ups.  Clearly many Hamiltonian systems including spin models such as
the quantum spin-glass \cite{Georges00}, or the disordered Heisenberg chains \cite{Brown08} are
potential places where such states can occur as eigenstates. The
number of particles allows to add complexity to the states in a
systematic manner, and interesting properties for entanglement unfold
in the process. Studies of entanglement in two-electron systems for instance
are found in \cite{Naudts07}. One other class of problems where there is 
potential to see the kind of transitions noted in this paper is in the study
of site-entanglement of fermions in a lattice \cite{Larsson06}, where the 
total number of qubits of this paper will be translated to the total number of
sites, the number of particles will be the number of fermions and the block
will refer to the sites within which the entanglement is found.  
It may also be noted that translationally invariant 
definite particle states with highly entangled nearest neighbor were constructed
as ``entangled rings'' \cite{WoottersRings}.

A previous study of entanglement in random one-particle states showed
that the averaged concurrence between any two qubits scales as $1/N$
\cite{arul03}.  Thus with increasing number of qubits entanglement
between any two still remains considerable, although decreasing, in
contrast to a full random state. In this paper it is shown that for
random two-particle states the average entanglement between two qubits
scales as $1/N^2$, while for three-particle states this becomes
exponentially small, as it goes as $\exp(-N \ln(N))$. Thus when the
number of particles exceeds two a transition is seen in the
entanglement between two qubits. It maybe noted that for full random
states it is not precisely known how such an entanglement scales with
the number of qubits. 

It is possible to generalize the results of concurrence between two
qubits to entanglement within the block $A$ having $m$ qubits of the
system for instance by studying the log-negativity measure
\cite{vidal02}.  Numerical and some analytical evidence points to the plausible result
that the entanglement decays with $N$ algebraically if the number of
particles in the subspace ($l$) is less than or equal to the
block-length ($m$). Once again the decay of
entanglement becomes exponential when the number of particles exceeds
the block-length. A study of the density of states of the reduced density matrix
also shows a transition when the number of particles exceeds the block size;
namely a divergence at the zero eigenvalue is replaced by a vanishing density.
This paper studies this divergence and how this impacts the partial transpose in such a way that entanglement undergoes the kind of transition that is discussed herein. 

The structure of the paper is as follows. In section ~\ref{sec:dps}, details of 
definite particle states and reduced density matrices of blocks of qubits
are given. 
In section ~\ref{sec:twoqbits} the entanglement between two
qubits is studied as a function of the number of particles present in
the state and as a function of the total number of qubits. Detailed
analytical results are derived utilizing concurrence as a measure of
entanglement. To demonstrate that the transition is independent of the 
particular measure of entanglement, as well as to facilitate generalization to larger
blocks, the log-negativity between two qubits is also considered here.
   In section ~\ref{sec:logneg} entanglement among larger
sets of qubits is studied by means of log-negativity and a few
analytical and several numerical results are presented. In
section~\ref{sec:dos} the density of states of the density matrix is
studied and it is shown that there is a transition in its character as
the number of particles exceeds the block size. Evidence is presented
that this results in, or is reflected as, the entanglement transition.

\section{Definite particle states}
\label{sec:dps}

A definite $l$-particle state is best written by grouping states with
a given number of particles present in one block, say $A$, and its
complementary block, say $B$. Let the number of qubits in block $A$ be
$m$ and let $l \geq m$. Label the states by the number of particles
(or total spin $S_z$) in subsets $A$ and $B$ to write:
\begin{eqnarray}
\label{eq:rand-state}
|\psi\rangle &= &  \sum_{j=1}^{{N-m \choose l}}c_{1j}^{(0)}|0\rangle_A |l\rangle^{j}_B 
+ \sum_{j=1}^{{N-m \choose l-1}} \sum_{i=1}^{{m \choose 1}} c_{ij}^{(1)} |1\rangle_A^{i} |l-1\rangle_B^j \nonumber\\
&+& \ldots + \sum_{j=1}^{{N- m \choose l-m}} c_{1j}^{(m)} |m\rangle_A |l-m\rangle_B^{j}.
\end{eqnarray}

The reduced density matrix of the subsystem $A$ denoted $\rho_A$, is the state of the block of qubits we are interested in
studying. These blocks
correspond to having a given number of particles, $k$, in the
subsystem $A$ and can be identified with one of the terms in the
expression for the state. Further, each of these blocks can be written
as $G_k = Q_k Q_k^\dagger$ where $Q_k$ is a matrix whose entries are
the coefficients $c_{ij}^{(k)}$ of the state. The dimensions (\#rows $\times$ \#columns) of, the in general rectangular matrices,  $Q_k$ and the square matrices $G_k$ are given by 
\begin{equation}\label{Qkdim}
\mbox{dim} \;Q_k = {m \choose k} \times {N-m \choose l-k},\;\; \mbox{dim} \; G_k = {m \choose k} \times {m \choose k}.
\end{equation}
 The condition that the trace
of a density matrix is unity implies that $\sum_k {\rm Tr}(Q_k
Q_k^\dagger) = 1.$ To construct the ensemble of $l$-particle states,
draw all the $\mathcal{N}={N \choose l}$ coefficients $c_{ij}^{(k)}$
from the normal distribution $N(0,1)$ and normalize them so that the
trace condition is met. This is equivalent to choosing them uniformly
with the only constraint being normalization \cite{wootters90}.

The case when the number of particles $l$ is less than the block length $m$
can be similarly written:
\begin{eqnarray}
\label{eq:rand-state.llessm}
|\psi\rangle &= &  \sum_{j=1}^{{N-m \choose l}}c_{1j}^{(0)}|0\rangle_A |l\rangle^{j}_B 
+ \sum_{j=1}^{{N-m \choose l-1}} \sum_{i=1}^{{m \choose 1}} c_{ij}^{(1)} |1\rangle_A^{i} |l-1\rangle_B^j \nonumber\\
&+& \ldots + \sum_{i=1}^{{m \choose l}} c_{i1}^{(l)} |l\rangle_A^i  |0\rangle_B.
\end{eqnarray}
In this case the last non-zero block has dimension ${m \choose l}$, however it has only one non-zero eigenvalue, as the corresponding density matrix $Q_l Q_l^{\dagger}$ is that of a unnormalized pure state. The sum of the dimensionality of the $G_k$ block matrices in this
case is less than $2^m$ and hence there are exact zero eigenvalues, in other words, the density matrix is rank-deficient.
For example in the simple case of one-particle states, the two qubit density matrix has one exact zero eigenvalue. In fact
it is easy to see that in general for one-particle states, the number of non-zero eigenvalues of the reduced density matrix of any number of qubits is at most two. From the dimension of the $Q_k$ matrices above we can formally write the minimum number of exact zero eigenvalues of the reduced density matrix for the case when $l <m$. Written in terms of rank:
\begin{equation}
\label{rank}
\mbox{rank}\; \rho_A \le \sum_{k=0}^{l} \mbox{min} \; \left [ {m \choose k}, \, {N-m \choose l-k} \right] < 2^m.
\end{equation}

In the case of $l \ge m$ the rank of typical reduced density matrices is full, that is the rank is $2^m$. This is seen by examining the
dimensions of each of the blocks $G_k$. From Eq.~(\ref{Qkdim}) it follows that typical $G_k$ will not  have exact zero eigenvalues if 
\begin{equation}
\label{mchoosek}
{m \choose k} \le {N-m \choose l-k}.
\end{equation}
A straightforward but nontrivial calculation shows that this is always the case if $N\ge 2l$. The equality is possible only when $l=m$. It is indeed assumed that
$N\ge 2l$ throughout this paper without any loss of generality, due to the particle-hole symmetry. In the special case
of $N=2l$ and $l=m$ all of the $Q_k$ matrices are of square type and the $G_k$ are formed then
from symmetric states. This has implications for the density of states as will be discussed later in this paper (see section ~\ref{sec:dos}). With these details about the structure of the reduced density matrices
of definite particle states, we now turn to a study of entanglement in
them. 

\section{Entanglement between two qubits in a definite particle state}
\label{sec:twoqbits}
The reduced density matrix of a block $A$ with $m=2$ qubits in a pure
state of $l \ge 2$ particles with the total number of qubits equal to $N$
can be written as:
\begin{equation}
\label{rhoA}
\rho_{A} = \left( \begin{matrix}
          a_{00} & 0 & 0 & 0\\
          0 & a_{11} & a_{12} & 0\\
          0 & a_{12}^* & a_{22} & 0\\         
          0 & 0 & 0  & a_{33}\\
\end{matrix}\right),
\end{equation}
where,
$a_{00} = \sum_{i = 1}^{\mu_0}(c_{1i}^{(0)})^2, \;\;\;a_{33} = \sum_{i = 1}^{\mu_2}(c_{1i}^{(2)})^2$, and 
\begin{equation} \left( \begin{matrix}
          a_{11} & a_{12}\\
          a_{12}^* & a_{22}\\         
\end{matrix}\right) = Q_1 Q_1^{\dagger}, \;
Q_1= \left( \begin{matrix}
         c_{11}^{(1)} & \ldots & c_{1\mu_1}^{(1)}\\
         c_{21}^{(1)} & \ldots & c_{2\mu_2}^{(1)}\\
\end{matrix}\right).
\label{QQ}
\end{equation}
Here $\mu_{i} ={N-2 \choose l-i}, \; i=0,1,2$.
The results presented in this work deal with real coefficients, ${c_{ij}^{(k)}}^* = c_{ij}^{(k)}$, a situation that would be relevant for example
for systems with time reversal symmetry. The central features, including the scaling, remain the same in the complex case. Also note that while the above expressions have been written when $l \ge 2$, it is straightforward to write the same when $l=1$, the case of 1-particle states. As a previous work
has dealt with 1-particle states \cite{arul03}, this is not considered further, however a detailed analysis of 
log-negativity in this case is presented later on in this paper. 

\subsection{Concurrence between two qubits}

Concurrence \cite{wootters98} is a measure of entanglement present between two qubits such as those in the subsystem $A$. The above structure in Eq.~(\ref{rhoA}), greatly simplifies the expression for concurrence \cite{connor01}
\begin{equation}
C =2 \; \mbox{max}(|a_{12}| - \sqrt{a_{00}a_{33}},0),
\end{equation}
and this allows for analytical estimates to be made, in contrast to 
the case of a full random state.
Due to the large number of coefficients $c_{ij}^{(k)}$ involved, it is a good approximation to assume that  the normalization constraint is only important to set their scale and that they are otherwise independent. This implies that these are {\it i.i.d.} random variables
drawn from the normal distribution $N(0,1/\mathcal{N})$; recall that $\mathcal{N}={N \choose l}$. The random numbers are ingredients
for the random variables $a_{00}$, $a_{12}$ and $a_{33}$ that determine the concurrence
and hence the entanglement.

The approach to finding the mean concurrence will be to first estimate the probability that it will be nonzero. The term $a_{12}$ involves a correlation
between two strings of normally distributed numbers, each of length  $\mu_1$, the two rows of the
matrix $Q_1$ in Eq.~(\ref{QQ}).
 On the other hand $a_{00}$ maybe taken to be effectively
its average and considered to be non-fluctuating, as it is a sum over $\sim N^l$ random terms.
 The following approximation then ensues:
\begin{equation}
\label{approx1}
\mbox{Pr}(C>0) \approx \mbox{Pr}( |a_{12}| - \sqrt{\langle a_{00} \rangle} \sqrt{ a_{33}} > 0).
\end{equation}

The distribution of $|a_{12}|$, $P_{12}$, is of central importance and can  obtained from, for example, from the probability density function of one of the marginals of the Wishart distribution for correlation matrices \cite{Wishart}. Suppressing the calculation, the result is 
\begin{equation}
P_{12}\left(  |a_{12}|= x \right) = 2 \mathcal{N} \frac{ K_{\nu}(\mathcal{N} x)}
{\sqrt{\pi}  \Gamma(\mu_1/2)}\left(\frac{\mathcal{N} x}{2}\right)^{\nu},
\end{equation}
where $K_\nu(x)$ is the modified Bessel function of the second kind,
and $\nu=(\mu_1-1)/2$.
The distribution of $\sqrt{a_{33}}$, $P_{33}$, follows from that of (square root of) a chi-squared
distribution with $\mu_2$ degrees of freedom:
\begin{equation}
P_{33}\left(  \sqrt{a_{33}}= y\right)=\dfrac{\mathcal{N}^{\mu_2/2}}{2^{\mu_2/2-1} \Gamma(\mu_2/2)} y^{\mu_2-1} e^{-\mathcal{N} y^2/2}.
\end{equation}

\begin{figure}
\includegraphics[width=0.8\linewidth,clip]{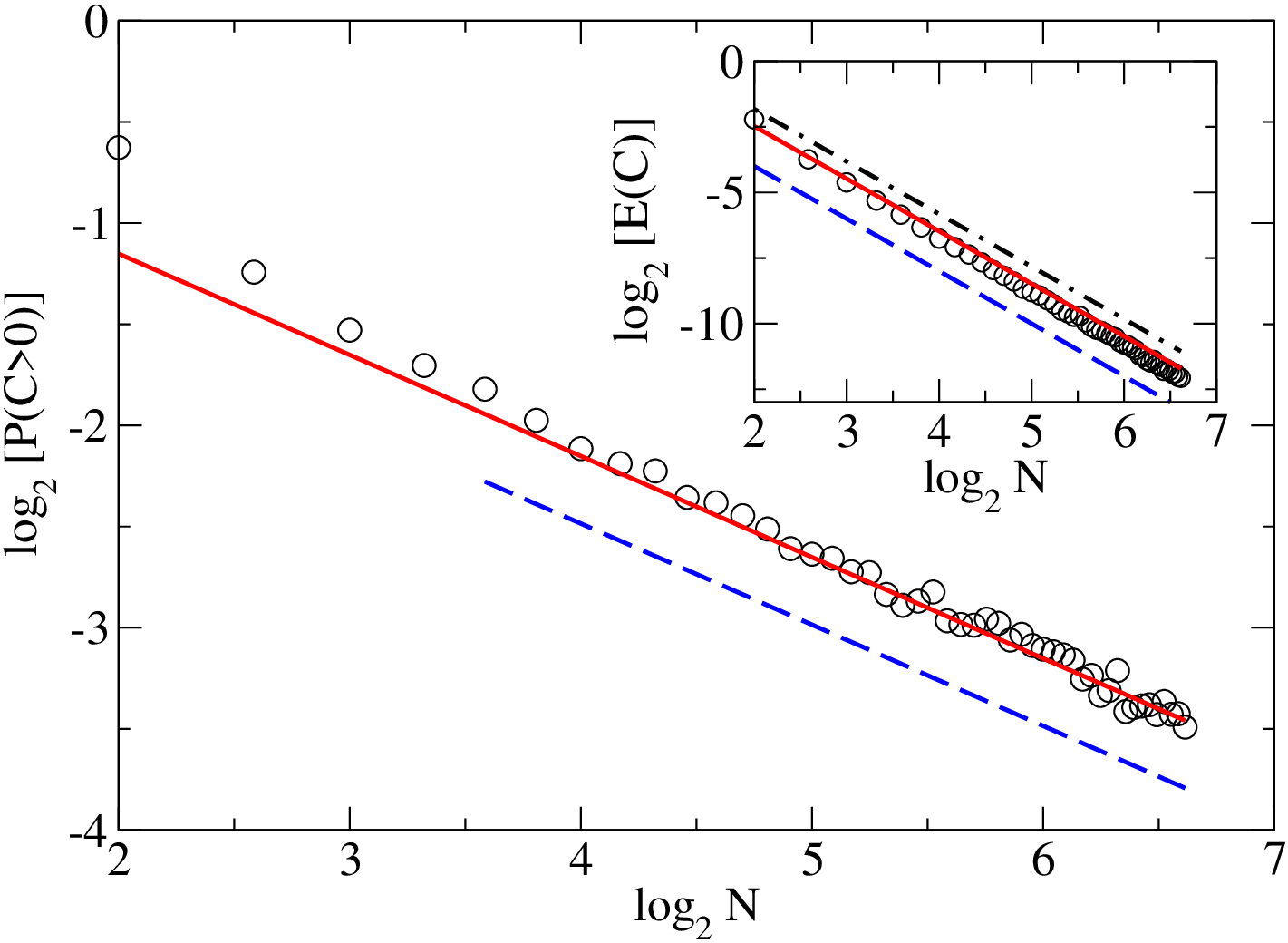}
\caption{The scaling of $\mbox{Pr}(C>0)$
for random two particle states with $N$ qubits. The dashed line is of slope $-1/2$, the circles are from numerical simulations, while the solid line
 is the estimate in Eq.~(\ref{2particle_upper_bound}).
Inset shows the average concurrence, the dashed line is of slope $-2$,
the dashed-dot line is the upper-bound while the solid line is the estimate in Eq.~(\ref{2particle_avg_ub}).}
\label{fig1}
\end{figure}

Thus in view of the approximation above it follows that:
\begin{equation}
\label{prob}
\mbox{Pr}(C>0)= \int_{0}^{\infty} P_{33}(y) \int_{\sqrt{\langle a_{00} \rangle } y }^{\infty} P_{12}(x) \,dx \, dy.
\end{equation}
This can be evaluated by steps which are outlined here: (a) change variable 
$y$ to $\sqrt{2 y/\mathcal{N}}$, and $\mathcal{N} x$ to $x$;
(b) use the integral representation $K_{\nu}(x)= \int_{0}^{\infty} e^{-x \cosh t } \cosh(\nu t) \, dt$, and change variable from $x \cosh t $ to $x$.  This leads to the exact expression (given the approximation in
Eq.~(\ref{approx1})) 
\begin{equation}
\mbox{Pr}(C>0)= \beta \int_{0}^{\infty} \dfrac{\cosh(\nu t)}{\cosh^{\nu+1} t}  \int_0^{\infty} y^{\frac{\mu_2}{2}-1}e^{-y} \,\Gamma\left( \nu+1,  \sqrt{\gamma y} \cosh t \right) \, dy\, dt.
\end{equation}
Here $\beta=2^{-\nu+1}/\sqrt{\pi} \Gamma(\mu_1/2) \Gamma(\mu_2/2)$, and $\gamma= 2 \langle a_{00} \rangle \mathcal{N}$. 
While further simplification is possible, for example by expanding $e^{-y}$, it is
expedient to seek a non-trivial upper bound that reveals the nature of the
decay with $N$, the number of qubits. A careful examination of the integrands indicate that this can be most easily achieved by using $e^{-y}<1$ and thus effectively removing the exponential
from the integral. The remaining integrals can be done exactly to
give the first inequality below, while the second follows from
standard inequalities for ratios of gamma functions:
\begin{equation}
\label{prob_upper_bound}
\mbox{Pr}(C>0) < \dfrac{2^{\mu_2}}{\gamma^{\mu_2/2} \sqrt{\pi}} \dfrac{\Gamma\left(\frac{\mu_1+\mu_2}{2} \right)}{\Gamma\left(\frac{\mu_1}{2}\right)}
\dfrac{\Gamma(\frac{\mu_2+1}{2})}{\Gamma(\frac{\mu_2}{2}+1)}\\
<\frac{1}{\sqrt{\pi}}\left(\dfrac{2 \mu_1 \eta}{\gamma}\right)^{\frac{\mu_2}{2}}   \dfrac{1}{\sqrt{\frac{\mu_2}{2}+\frac{1}{4}}}, 
\end{equation}
where $\eta=1+(\mu_2-2)/(2 \mu_1)$. Note that when 
the number of particles is much less than the number of qubits, $\eta \approx 1$.
For the case of two particle states, $l=2$, the first inequality yields
\begin{equation}
\label{2particle_upper_bound}
\mbox{Pr}(C>0) <\frac{2 \sqrt{2}}{\pi} \frac{1}{\sqrt{N}},
\end{equation}
as $\mu_1=N-2$ and $\mu_2=1$. The inequality is valid for large $N$,
especially as the value of $\langle a_{00} \rangle$ is taken to be 1.
As a matter of fact that this is an excellent estimate by itself is
seen in Fig.~(\ref{fig1}).

The two particle case is of special interest and can be essentially
derived from simpler formulae, if it is observed that the fluctuations
in $a_{33}$, arising from a single realization of the random
variables, is more than the others. Note that: $a_{00} \sim
\mu_0/\mathcal{N} \sim 1$, $a_{33} \sim \mu_2/\mathcal{N} \sim 1/N^2$,
and $|a_{12}|^2 \sim \mu_1/\mathcal{N}^2 \sim 4/N^3$. Hence typically
the concurrence will indeed be zero. Replacing the average values for
the fluctuating $a_{00}$ and $|a_{12}|$ results in
$\mbox{Pr}(C > 0) \approx \mbox{Pr}( \sqrt{a_{33}} < \sqrt{\frac{2}{\pi}}\frac{2}{N^{3/2}}) =\frac{2 \sqrt{2}}{\pi} \frac{1}{\sqrt{N}},$
coinciding with the upper bound just derived.
The average value of $|a_{12}|$ is used, rather than the (square root of the) average of  $|a_{12}|^2$; the exact distribution can be used to show that  $\langle |a_{12}|^2 \rangle = \frac{\pi}{2}(\langle |a_{12}| \rangle ^2)$. Thus for two particle states the probability of concurrence being positive decreases algebraically, in contrast to the one-particle case when $P(C>0)=1$, as $a_{33}=0$.

\begin{figure}
\includegraphics[width=0.8\linewidth,clip]{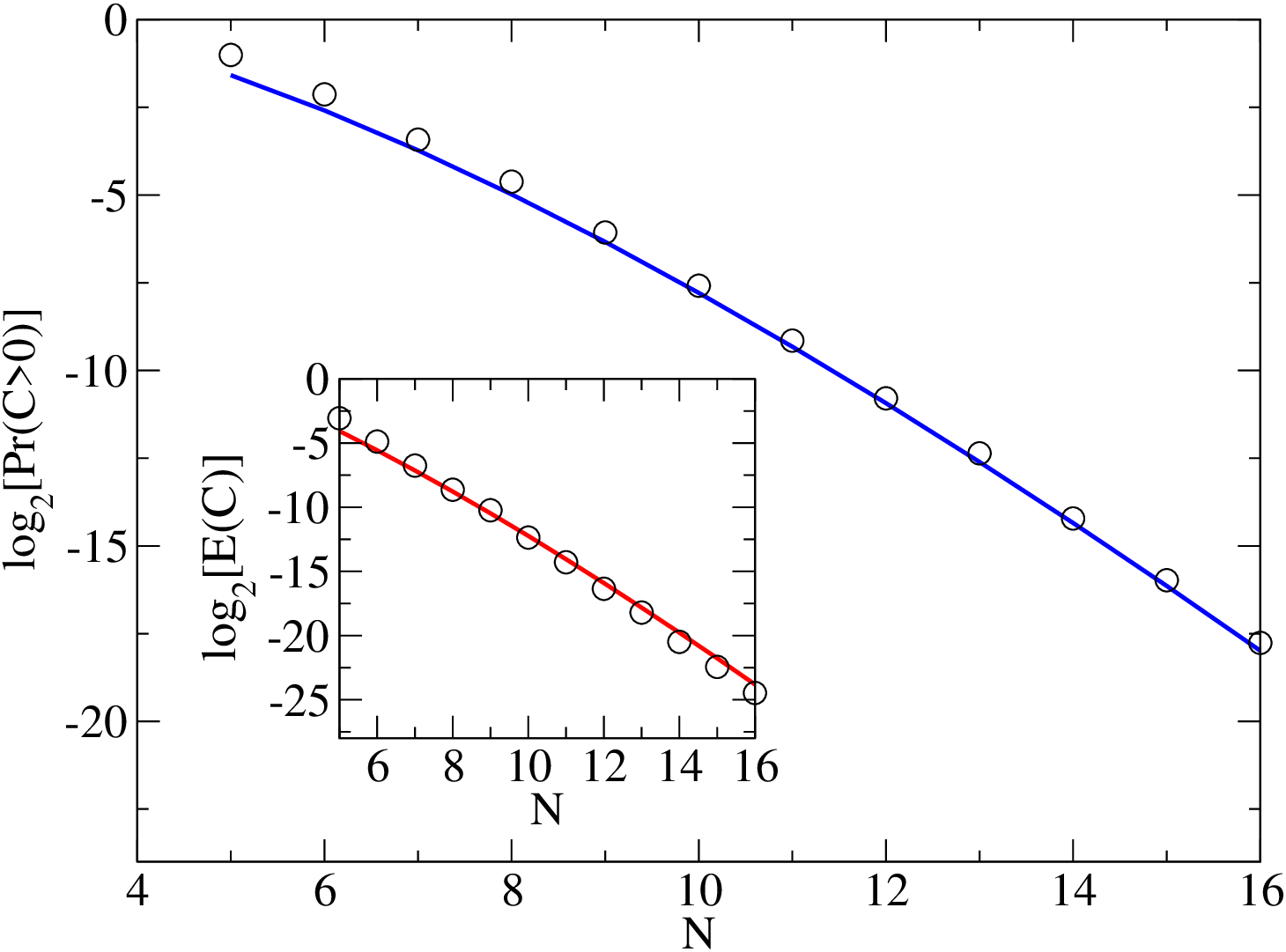}
\caption{ The probability $ \mbox{Pr}(C>0)$ for three particle states,
and the average concurrence as number of qubits $N$ is changed.
Note that the y-axes are on a logarithmic scale. The circles are from 
numerical simulations while the solid line in the case of $\mbox{Pr}(C>0)$ is from an exact numerical evaluation of Eq.~(\ref{prob}).}
\label{fig2}
\end{figure}

For $l=3$, three particle states, a completely different behavior is obtained as $\mu_1 \sim N^2/2$, $\mu_2\sim N$, and $\gamma \sim N^3/3$ which result in
\begin{equation}
\label{3partprob}
\mbox{Pr}(C>0) < \sqrt{\frac{2}{\pi\, N}} \exp\left(-\frac{N}{2} \log (N/3)\right).
\end{equation}
Unlike the two-particle case the probability that the concurrence is positive decreases at least exponentially with the number of qubits,
see Fig.~(\ref{fig2}).
Another new feature is that it is quite essential to take into account 
the fluctuations in {\it both} $|a_{12}|$ and in $a_{33}$. Ignoring say the fluctuations in $a_{33}$ results in much smaller estimates of
the probability than what is found.

When $l>2$, but still much less than $N$, the upper-bound in Eq.~(\ref{prob_upper_bound}) does not estimate the probability accurately. While it can be made tighter, this is indeed a good bound as it is simple, decreases with $N$, and shows the advertized transition in the entanglement as one particle is added to a two particle state. It will be 
seen that the entanglement hitherto shared between two qubits will
now be available for three-body and multi-party entanglement. 

If the fraction of particles $p=l/N$ is of order 1 (and less than $1/2$), the states are ``macroscopically" occupied; employing the approximation that ${N \choose Np}\sim e^{SN}$ where $S=-p \ln(p) -(1-p)\ln(1-p)$ is the binary entropy corresponding to probability $p$,  results in the upper bound $
\mbox{Pr}(C>0) <d_{1}e^{-SN/2} \, e^{-d_2e^{SN}}$,
where $d_1$ and $d_2$ are positive constants of order 1. However the upper-bound in Eq.~(\ref{prob_upper_bound}) has to be used with caution as it can be rendered trivial if $(2\mu_1\eta/\gamma)>1$, and consequently $d_2$ becomes negative. Thus for $N=10$ qubits and $l=5$ particles the upper-bound $\approx 2.2$ is trivial while for $l=4$ it is $\approx 1.5 \times 10^{-5}$. While $N=12$, $l=6$ results in a trivial bound, $l=5$ results in $\mbox{Pr}(C>0)<3.3 \times 10^{-15}$. Similarly 
when $N=14$ and $l=6$, the upper-bound is $\approx 1.6 \times 10^{-43}$, it is improbable that two qubits will be entangled.

The mean concurrence, $\mathbb{E}(C)$ is now estimated. In the two particle case for instance 
 \begin{equation}
 \label{2particle_avg_ub}
 \mathbb{E}(C)\sim 2 \langle |a_{12}|\rangle \mbox{Pr}(C>0) \sim \dfrac{16}{\pi^{3/2} N^2}.
 \end{equation}
  A more general estimate is possible as 
$\mathbb{E}(C)=\mathbb{E}[2(x-\sqrt{\langle a_{00} \rangle }y) \Theta(x-\sqrt{\langle a_{00} \rangle }y)]$. Using the distribution $P_{12}(x)P_{33}(y)$ and following the same steps as outlined for the probability
above it follows that 
\begin{equation} \mathbb{E}(C) <
\dfrac{2^{\mu_2+2}}{\mathcal{N} \gamma^{\mu_2/2} \sqrt{\pi}} \dfrac{\Gamma\left(\frac{\mu_1+\mu_2+1}{2} \right)}{\Gamma\left(\frac{\mu_1}{2}\right)}
<\frac{2 \sqrt{\gamma}}{\mathcal{N}\sqrt{\pi}}\left(\dfrac{2 \mu_1 \eta'}{\gamma}\right)^{\frac{\mu_2+1}{2}}  
\end{equation}
where $\eta'=1+(\mu_2-1)/(2 \mu_1)$. In the two particle case this gives $\mathbb{E}(C)<8/\sqrt{\pi} N^2$, which is quite close to the estimate above. The exponential decay for three or more particles is manifest. The mean concurrences are shown in the insets of Figs.~(\ref{fig1}),(\ref{fig2}).

\subsection{Log-negativity among two qubits}

The vanishingly small two qubit entanglement for more than two-particle states ($l=2$) goes into multiparty entanglement. A measure of entanglement that can be easily extended to a subsystem having more than two qubits is the log-negativity~\cite{vidal02} and is given by $E_{LN}(\rho^{AB}) = \log(||\rho_{AB}^{\Gamma}||),$
where $||\rho^{\Gamma}||$ is the trace norm of the partial transpose matrix $\rho^{\Gamma}$~\cite{peres96}. 
When log-negativity is zero the state is said to have  positive partial transpose (PPT) and in that case it is either separable or bound entangled
\cite{mhorodeckibound}. When log-negativity is greater than zero the state is said to have negative partial transpose (NPT) and in that case it is entangled.

On studying entanglement in a  block length of $2$ we get the entanglement between two qubits as measured by log-negativity. This decays algebraically as $1/N^{3.5}$ in contrast to the $1/N^2$ behavior of the concurrence for the case of two particles, but becomes exponential when the particle number is increased to three or more. See Fig.~(\ref{2qubitfig15}) for details.
 Thus on using a different measure of entanglement while indeed the 
exponents change the qualitative nature of the decay with the number of particles remains intact. 
It is also useful to contrast the case of 1-particle states and therefore log-negativity is now derived between
any two qubits for both 1- and 2-particle states.

\subsubsection{Block of 2 qubits and 1-particle states}
In this case the reduced density matrix is block diagonal consisting of two square block and is given as follows:

\begin{equation}
\label{rhoALN}
\rho_{A} = \left( \begin{matrix}
          a_{00} & 0 & 0 & 0\\
          0 & a_{11} & a_{12} & 0\\
          0 & a_{12}^* & a_{22} & 0\\         
          0 & 0 & 0  & 0\\
\end{matrix}\right),
\end{equation}
where,
$a_{00} = \sum_{i = 1}^{N-2}(c_{1i}^{(0)})^2$, and 
\begin{equation} \left( \begin{matrix}
          a_{11} & a_{12}\\
          a_{12}^* & a_{22}\\         
\end{matrix}\right) = Q_1 Q_1^{\dagger}, \mbox{and} \;
Q_1= \left( \begin{matrix}
         c_{11}^{(1)} \\
         c_{21}^{(1)} \\
\end{matrix}\right),
\end{equation}
where to remind the reader the coefficients $c_{ij}$ are as defined in Eq.~(\ref{eq:rand-state}).

 Partial transpose (PT) on the second qubit of $\rho_{A}$ results in 
\begin{equation}
\rho_{A}^{\Gamma}  = \left( \begin{matrix}
          a_{00} & 0 & 0 & a_{12}\\
          0 & a_{11} & 0 & 0\\
          0 & 0 & a_{22} & 0\\         
          a_{12}^* & 0 & 0  & 0\\
\end{matrix}\right).
\end{equation}
The eigenvalues of $\rho_{A}^{\Gamma}$ (as always in this paper, for the case of real coefficients  ${c_{ij}^{(k)}}$)
are $\Lambda_{\pm}=(a_{00} \pm \sqrt{(a_{00})^2 +4a_{12}^2})/2$,  $a_{11}$ and $a_{22}$. 
The only negative eigenvalue is $\Lambda_{-}$.
From the assumptions of randomness of the state, for one-particle states we 
see that $a_{00} \sim 1$, $a_{12}^2 \sim 1/N^2$. Using this the negative eigenvalue can be approximated by 
$-a_{12}^2/a_{00}\; \sim \; -a_{12}^2$.
The log-negativity is given by $E_{LN}=\log(1-2 \sum_{i} \omega_i) \approx -2 \sum_{i} \omega_i \approx 2 a_{12}^2 \approx 2/N^2$,
where the sum is over all the negative eigenvalues ($\omega_i$) of $\rho_{A}^{\Gamma}$ and the approximation holds good since the 
$\omega_i$'s are much smaller than $1$. Indeed one finds that this estimate is in good agreement with the numerical results shown in Fig.~(\ref{2qubitfig15}). Note that the average concurrence between any two qubits for 1-particle states scales as $1/N$ \cite{arul03}.

\subsubsection{Block of 2 qubits and 2-particle states}

In this case reduced density matrix is block diagonal consisting of three square block and is given as in Eq.~(\ref{rhoA}). Partial transpose on the second qubit of $\rho_{A}$ results in 
\begin{equation}
\rho_{A}^{\Gamma}  = \left( \begin{matrix}
          a_{00} & 0 & 0 & a_{12}\\
          0 & a_{11} & 0 & 0\\
          0 & 0 & a_{22} & 0\\         
          a_{12}^* & 0 & 0 & a_{33}\\
\end{matrix}\right).
\end{equation}

\begin{figure}
\includegraphics[width=0.49\linewidth,clip]{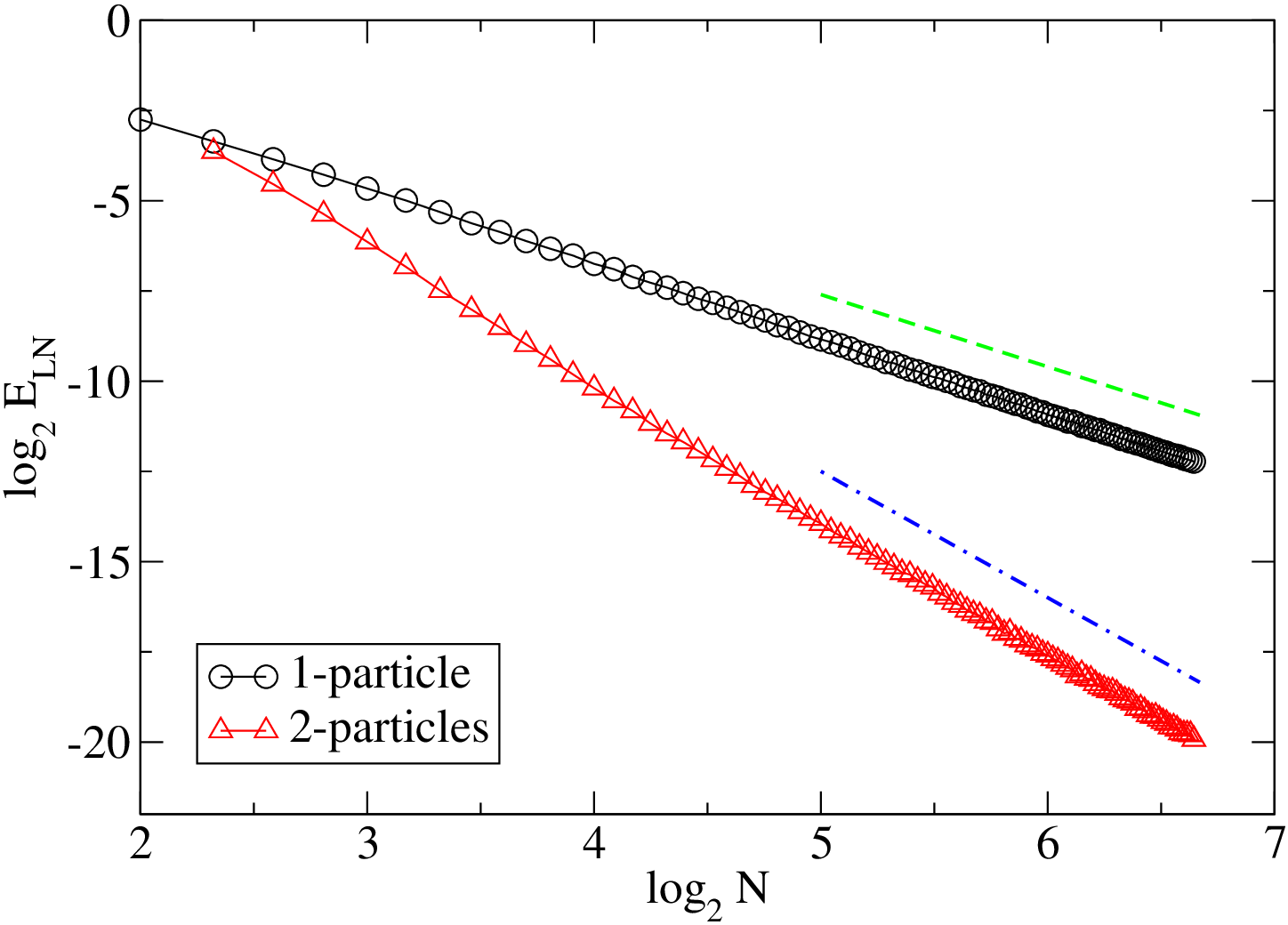}
\includegraphics[width=0.49\linewidth,clip]{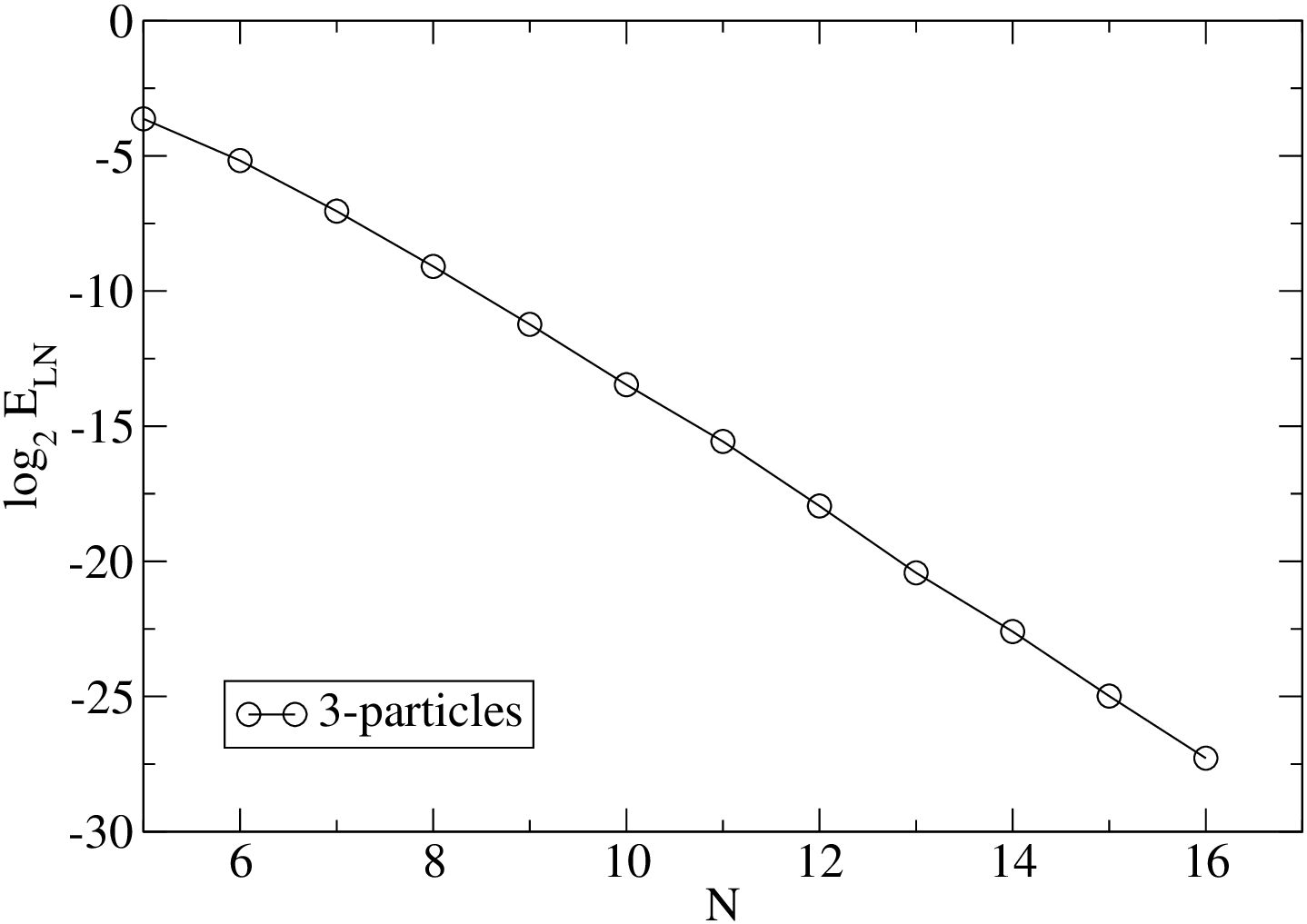}  
\caption{Scaling of log-negativity ($E_{LN}$) in a block of 2 qubits, with the total number $N$ of qubits for (left) one, two-particle and 
(right) three-particle states. In figure on left dashed line, dashed-dot line are having slope of -2, -3.5 respectively.}
\label{2qubitfig15}
\end{figure}

The eigenvalues of this are 
\[
\Lambda_{\pm}=\dfrac{1}{2} \left( a_{00}+a_{33} \pm \sqrt{(a_{00}+a_{33})^2-4(a_{00}a_{33}-a_{12}^2)} \right)  
\]
and the pair of $a_{11}$ and $a_{22}$. Again the only possible negative eigenvalue is $\Lambda_{-}$ which occurs
when $a_{00}a_{33}-a_{12}^2$ is negative. In the case of two-particle sates, as we see earlier,  $a_{00} \sim 1$, $a_{12}^2 \sim 4/N^3$ and $a_{33} \sim 1/N^2$ and thus  the eigenvalue $\Lambda_{-}$ can be approximated by 
$(a_{00}a_{33}-a_{12}^2)/(a_{00}+a_{33})\; \sim \; a_{00}a_{33}-a_{12}^2$.
Thus $E_{LN} = 2 a_{12}^2 \; $Pr$((a_{00}a_{33}-a_{12}^2) < 0)$. Using Eq.~(\ref{2particle_upper_bound}) 
we find that
\begin{equation}
E_{LN} \sim \dfrac{16\sqrt{2}}{\pi} \dfrac{1}{N^{3.5}} \nonumber
\end{equation}
which is in good agreement with the numerical results  shown 
in Fig.~(\ref{2qubitfig15}). In the case of a block of 2 qubits and 3-particle states one finds that the log-negativity scales exponentially 
with the total number of qubits $N$ as shown in  Fig.~(\ref{2qubitfig15}). This follows on using the exponentially small probability
for the concurrence to be positive (see Eq. ~(\ref{3partprob})) and a similar analysis as above.

\section{Entanglement among larger subsets of qubits}
\label{sec:logneg}

\begin{figure}
\includegraphics[width=0.49\linewidth,clip]{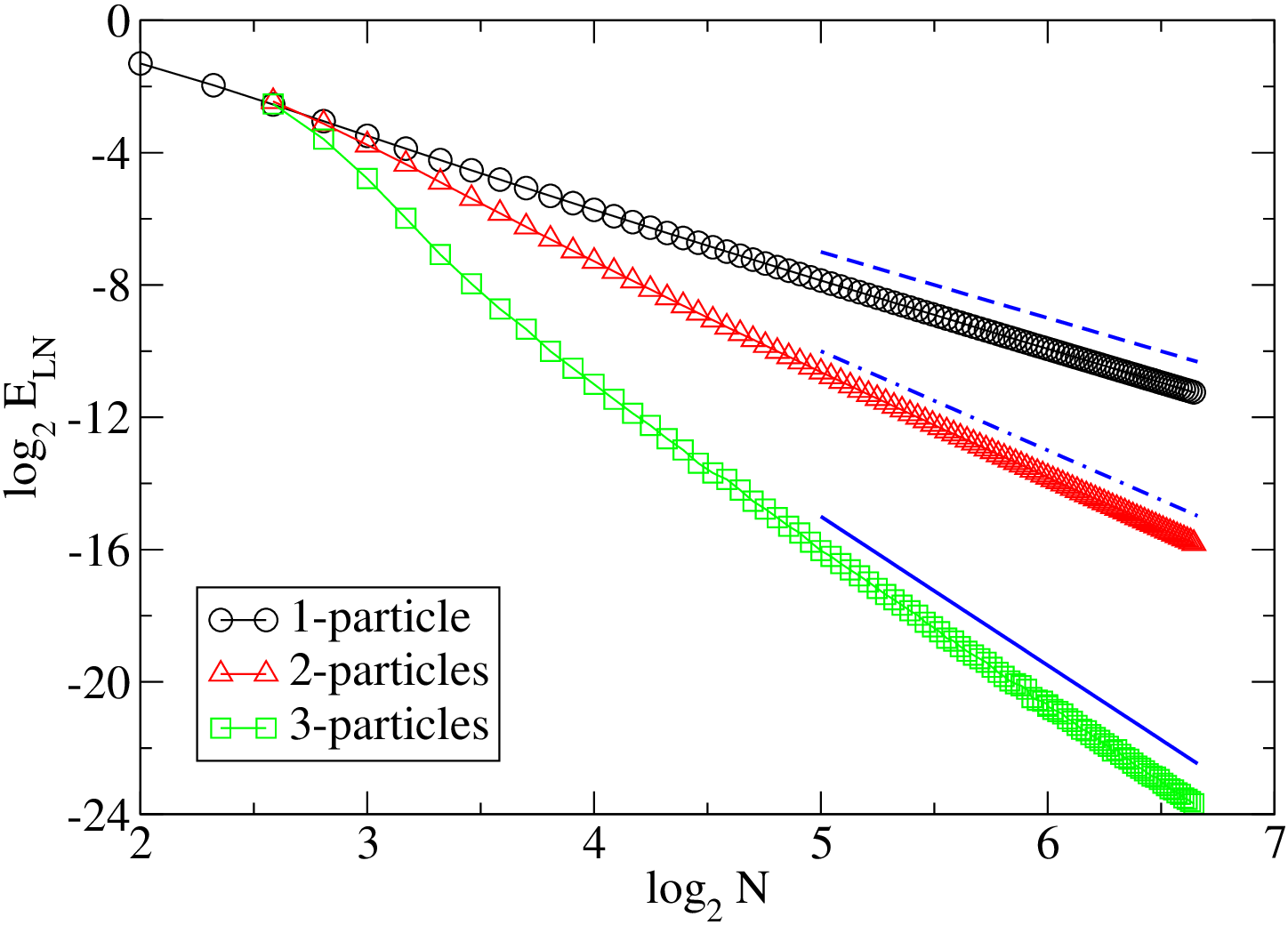}
\includegraphics[width=0.5\linewidth,clip]{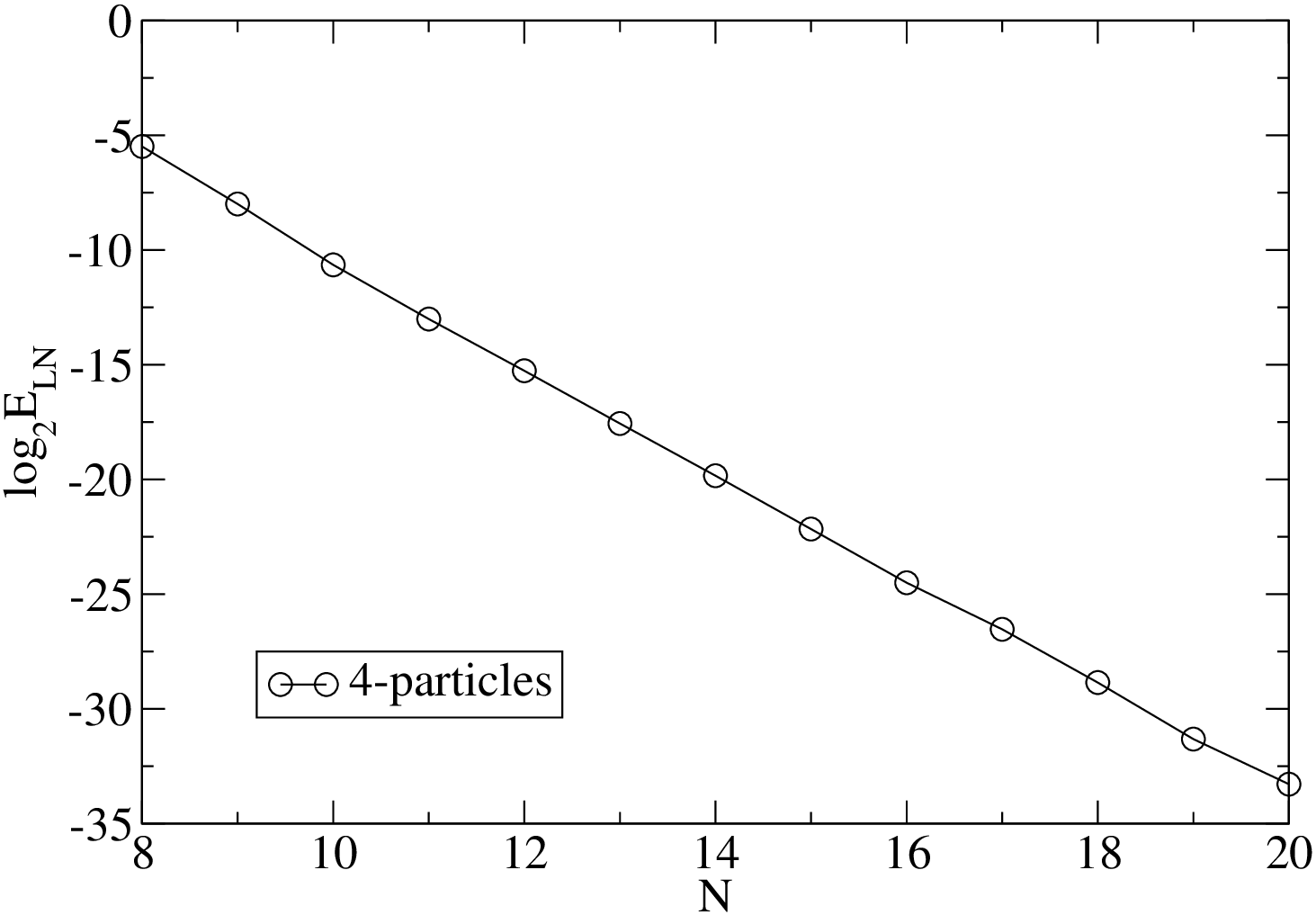}  
\caption{Scaling of log-negativity ($E_{LN}$) in a block of 3 qubits, with the total number $N$ of qubits for (left) one, two and 
three-particle and (right) four-particle states. In figure on left dashed line, dashed-dot line and simple straight line are 
having slope of -2, -3, -4.5 respectively.}
\label{fig10}
\end{figure}

While the previous section has dealt exclusively with a ``block" of two qubits, here we take
larger subsets of qubits to belong to block $A$. The Log-negativity measure will be used once again. 
 A transition similar to the one above is exhibited for the entanglement between a qubit and the other pair when a block of $3$ qubits is considered. Algebraic decay of the log-negativity for $l\le 3$ is replaced by exponential decay for $l>3$, see Fig.~(\ref{fig10}).
The decay with particle number is algebraic and the exponent is the ``slope" in Table \ref{table:block3}. 
Further results for block lengths of 4 are presented in Table \ref{table:block4}, in which case there are two distinct types of partitions, entanglement between two pairs of qubits (denoted as 2+2) and 
between a triple and a lone qubit (denoted 3+1).  However the numerics becomes considerably more 
difficult thereon, and the slopes given may not be entirely converged. However, the transition 
from algebraic to exponential is a robust feature.
\begin{table}[ht]
\caption{Block length 3.}
 \centering
 \begin{tabular}{|c| c|}
 \hline \hline
 Particle \# ($l$) & Decay with number of qubits $N$.\\
 \hline 
 1 & Alg.: slope = -2\\
 2 & Alg.:  slope = -3\\
 3 & Alg.:  slope = -4.5\\
 4 & Exponential\\
 \hline
 \end{tabular}
 \label{table:block3}
 \end{table}
 \begin{table}[ht]
\caption{Block length 4. Cases of (2+2,3+1)}
 \centering
 \begin{tabular}{|c| c|}
 \hline \hline
 Particle \# ($l$) & Decay with number of qubits $N$.\\
 \hline 
 1 & Alg.: slopes = (-2.1, -2.1)\\
 2 & Alg.: slopes = (-2.1, -3.1)\\
 3 & Alg.: slopes = (-4.1, -4.1)\\
 4 & Alg.: slopes = (-5.7, -5.7)\\
 5 & Exponential\\
 \hline
 \end{tabular}
 \label{table:block4}
 \end{table}

It is possible to extend the analysis of the log-negativity of two qubits to the case of a block of three qubits and derive some of the exponents as stated in the table. 
Now large $N$ formulae for log-negativity in the case of a block of 3 qubits in 1-particle and 2-particle states are derived. It is shown that the log-negativity decays as $4/N^2$, $16/N^3$ for these
cases respectively. 

\setcounter{subsubsection}{0}
\subsubsection{Block of 3 qubits and 1-particle case.}
In this case the reduced density matrix is block diagonal consisting of two square blocks, one of them ($a_{00}$) being just a number:
\begin{equation}
\label{rhoA2}
\rho_{A} = \left( \begin{matrix}
          a_{00} & 0_{1 \times 3} & 0_{1 \times 4} \\
          0_{3 \times 1} & Q_1 Q_1^{\dagger} & 0_{3 \times 3}\\
          0_{4 \times 1} & 0_{4 \times 3} & 0_{4 \times 4}\\         
\end{matrix}\right),
\end{equation}

where $a_{00} = \sum_{i = 1}^{N-3}(c_{1i}^{(0)})^2$, $0_{p \times q}$ are zero matrices with dimensions $p \times q$
 and 
\begin{equation}
Q_1 Q_1^{\dagger}=\left( \begin{matrix}
          a_{11} & a_{12} & a_{13}\\
          a_{12}^* & a_{22} & a_{23}\\   
	  a_{13}^* & a_{23}^* & a_{33}\\
\end{matrix}\right), \;
Q_1 = \left( \begin{matrix}
	  c_{11}^{(1)}\\
	  c_{21}^{(1)}\\
	  c_{31}^{(1)}\\
\end{matrix}\right).
\end{equation}
Being 1-particle states these have only two nonzero eigenvalues in general. On PT
it is seen that there are {\it four} nonzero eigenvalues.
Partial transpose on the third qubit of $\rho_{A}$ results in
\begin{equation}
\rho_{A}^{\Gamma} = \left( \begin{matrix}
          a_{00} & 0 & 0 & 0 &  a_{12} & a_{13} & 0 & 0\\
          0 & a_{11} & 0 & 0 & 0 & 0 & 0 & 0\\
          0 & 0 & a_{22} & a_{23} & 0 & 0 & 0 & 0\\         
          0 & 0 & a_{23}^* & a_{33}  & 0 & 0 & 0 & 0\\
	  a_{12}^* & 0 & 0 & 0 & 0 & 0 & 0 & 0\\
	  a_{13}^* & 0 & 0 & 0 & 0 & 0 & 0 & 0\\
	  0 & 0 & 0 & 0 & 0 & 0 & 0 & 0\\
	  0 & 0 & 0 & 0 & 0 & 0 & 0 & 0\\
\end{matrix}\right).
\end{equation}
The nonzero eigenvalues of $\rho^{\Gamma}$ are $a_{11}$, $a_{22}+a_{33}$ and  
\begin{equation}
\Lambda_{\pm}=\dfrac{1}{2} \left( a_{00} \pm \sqrt{a_{00}^2+ 4(a_{12}^2+a_{13}^2)}\right). \nonumber
\end{equation}
Note that there are correlations in the entries of the $Q_1 Q_1^{\dagger}$ matrices, such
as $a_{22} a_{33} -a_{23}^2=0$ as the state from the density matrix is constructed 
is an (unnormalized) pure state. 
Here $a_{00} \sim 1$, $a_{12}^2$ and $a_{13}^2 \sim 1/N^2$.
It can be seen that only one of the four nonzero eigenvalues is negative and it is 
$\Lambda_{-}$.
Using this, the negative eigenvalue can be approximated as
$-(a_{12}^2+a_{13}^2)/a_{00} \approx -(a_{12}^2+a_{13}^2)$.
The log-negativity is therefore given by $E_{LN} \approx -2 \sum_{i} \omega_i \approx 2(a_{12}^2+a_{13}^2) \approx 4/N^2$.
This estimate is in good agreement with numerical results as shown in Fig.~(\ref{fig10}).

\subsubsection{Block of 3 qubits and 2-particle case.}
In this case the reduced density matrix is block diagonal consisting of three square blocks and is given as follows:
\begin{equation}
\rho_{A} = \left( \begin{matrix}
          a_{00} & 0_{1 \times 3} & 0_{1 \times 3} & 0_{1 \times 1} \\
          0_{3 \times 1} & Q_1 Q_1^{\dagger} & 0_{3 \times 3} & 0_{3 \times 1}\\
          0_{3 \times 1} & 0_{3 \times 3} & Q_2 Q_2^{\dagger} & 0_{3 \times 1}\\         
	  0_{1 \times 1} & 0_{1 \times 3} & 0_{1 \times 3} & 0_{1 \times 1} \\         
\end{matrix}\right),
\end{equation}

where $a_{00} = \sum_{i = 1}^{\alpha_0}(c_{1i}^{(0)})^2$, and 
\begin{equation}
 Q_1 Q_1^{\dagger}=\left( \begin{matrix}
          a_{11} & a_{12} & a_{13}\\
          a_{12}^* & a_{22} & a_{23}\\   
	  a_{13}^* & a_{23}^* & a_{33}\\
\end{matrix}\right), \;
Q_1 = \left( \begin{matrix}
         c_{11}^{(1)} & \ldots & c_{1\alpha_1}^{(1)}\\
         c_{21}^{(1)}  & \ldots & c_{2\alpha_1^{(1)}}\\
	 c_{31}^{(1)}  & \ldots & c_{3\alpha_1^{(1)}}\\ 
\end{matrix}\right),
\end{equation}
\begin{equation}
Q_2 Q_2^{\dagger}=\left( \begin{matrix}
          a_{44} & a_{45} & a_{46} \\
          a_{45}^* & a_{55} & a_{56}\\   
	 a_{46}^* & a_{56}^* & a_{66}\\
\end{matrix}\right), \;
Q_2 = \left( \begin{matrix}
	  c_{11}^{(2)}\\
	  c_{21}^{(2)}\\
	  c_{31}^{(2)}\\
\end{matrix}\right).
\end{equation}
Here $\alpha_{i} ={N-3 \choose 2-i}, \; i=0,1,2$.
Partial transpose on the third qubit of $\rho_{A}$ results in 
\begin{equation}
\rho_{A}^{\Gamma} = \left( \begin{matrix}
          a_{00} & 0 & 0 & 0 &  a_{12} & a_{13} & 0 & 0\\
          0 & a_{11} & 0 & 0 & 0 & 0 & 0 & 0\\
          0 & 0 & a_{22} & a_{23} & 0 & 0 & 0 &  a_{46}\\         
          0 & 0 & a_{23}^* & a_{33}  & 0 & 0 & 0 &  a_{56}\\
	  a_{12}^* & 0 & 0 & 0 & a_{44} & a_{45} & 0 & 0\\
	  a_{13}^* & 0 & 0 & 0 & a_{45}^* & a_{55} & 0 & 0\\
	  0 & 0 & 0 & 0 & 0 & 0 & a_{66} & 0\\
	  0 & 0 &  a_{46}^* &  a_{46}^* & 0 & 0 & 0 & 0\\
\end{matrix}\right).
\end{equation}
In this case, while the density matrix has one zero eigenvalue, the partial transpose has
no zero eigenvalue. Apart from the eigenvalues $a_{11}$ and $a_{66}$, the other six 
eigenvalues are those of the matrices $A$ and $B$, where:
\begin{equation}
A = \left( \begin{matrix}
	  a_{00} & a_{12} & a_{13}\\
	  a_{12}^* & a_{44} & a_{45}\\
	  a_{13}^* & a_{45}^* & a_{55}\\
\end{matrix}\right),\;
B = \left( \begin{matrix}
	  a_{22} & a_{23} & a_{46}\\
	  a_{23}^* & a_{33} & a_{56}\\
	  a_{46}^* & a_{56}^* & 0\\
\end{matrix}\right).
\end{equation}
The characteristic equation of matrix $A$ is 
\begin{eqnarray}
 \lambda^3-\lambda^2(a_{00}+a_{44}+a_{55})-\lambda(a_{45}^2+a_{12}^2+a_{13}^2- a_{00}a_{44}-a_{00}a_{55}-a_{44}a_{55})\\ \nonumber
+(a_{00}a_{45}^2+a_{55}a_{12}^2+a_{44}a_{13}^2-a_{55}a_{00}a_{44}-2 a_{12}a_{13}a_{45}) &=& 0.
\end{eqnarray}
Here we see that average of coefficients of $\lambda^2$ and $\lambda$ goes as $-1-2/N^2$ and $4/N^2$ while that of constant term goes as 
$16/N^5$. Thus typically the determinants of $A$ and $B$, which are the negative of the constant term, are negative, and hence the three eigenvalues of each matrix can either all be negative
or have one lone negative value. That the latter is the case follows on noting that the traces of
these two matrices are positive.
 
 One may estimate the negative eigenvalue now. Assuming that the terms containing $\lambda^3$
 and $\lambda^2$ are of smaller order, and keeping only the linear and constant terms one gets
 that the negative eigenvalues of $A$ is approximately $-4/N^3$. It is immediately verified that
 the assumptions just made are justified. 
The characteristic equation of matrix $B$ is 
\[
  \lambda^3-\lambda^2(a_{22}+a_{33})-\lambda(a_{56}^2+a_{46}^2+a_{23}^2-a_{22}a_{33})
+a_{22}a_{56}^2+a_{33}a_{46}^2-2 a_{23}a_{46}a_{56} = 0.\]
The average of the coefficients of $\lambda^2$ and $\lambda$ go as $-2/N$ and $4/N^2$ while that of the constant term goes as $16/N^5$. Using an argument similar to that used to approximate the  negative eigenvalue of the matrix $A$, we can approximate the same for matrix $B$ as $-4/N^3$.
The log-negativity is given by $E_{LN} \approx -2 \sum_{i} \omega_i \approx 8/N^3+8/N^3 = 16/N^3$.
This estimate is again in very good agreement with numerical results as shown in Fig.~(\ref{fig10}), including both the exponent and
the constant.

\section{Density of states before and after PT}
\label{sec:dos}

\begin{figure}
\includegraphics[width=0.49\linewidth,clip]{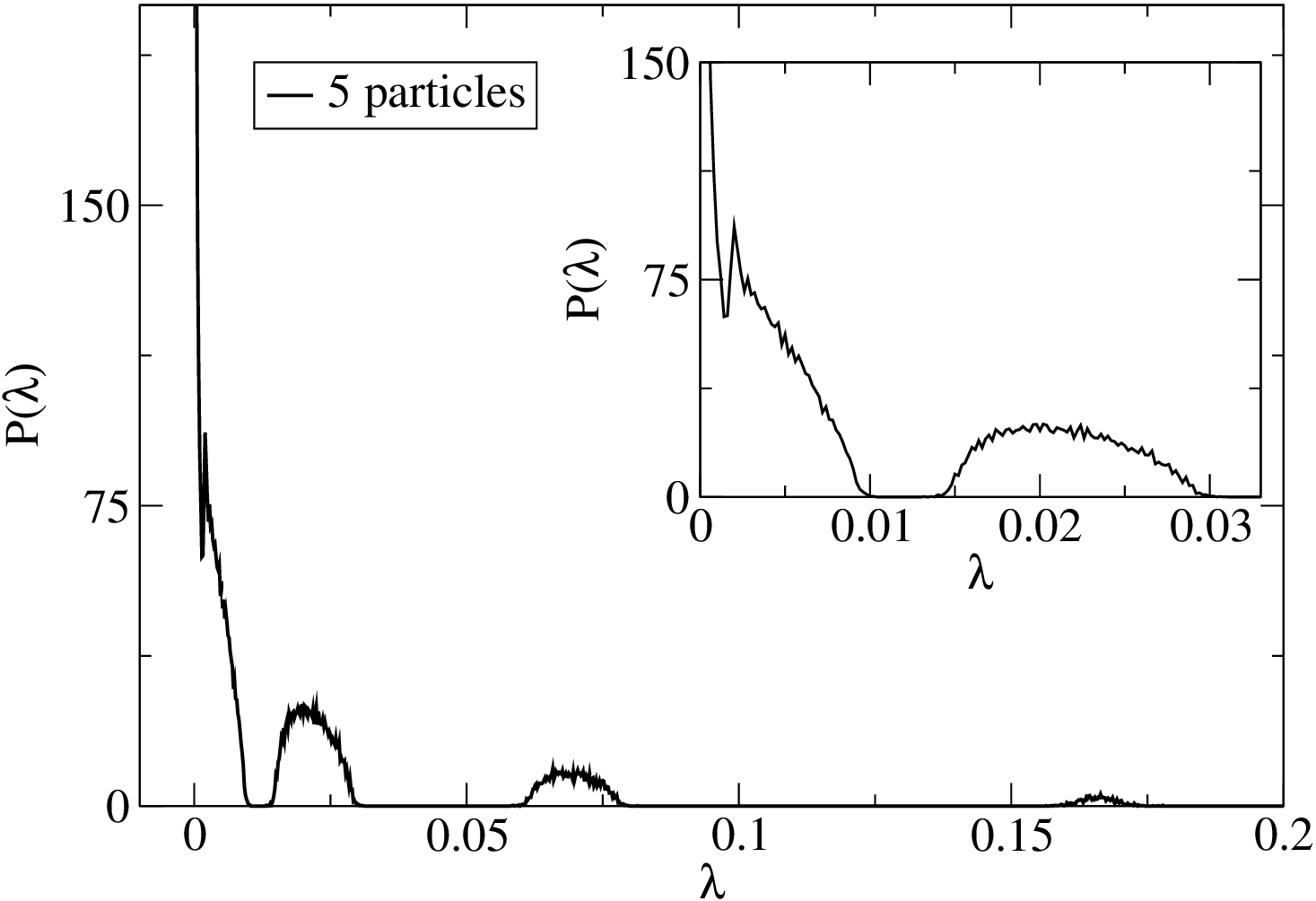}
\includegraphics[width=0.49\linewidth,clip]{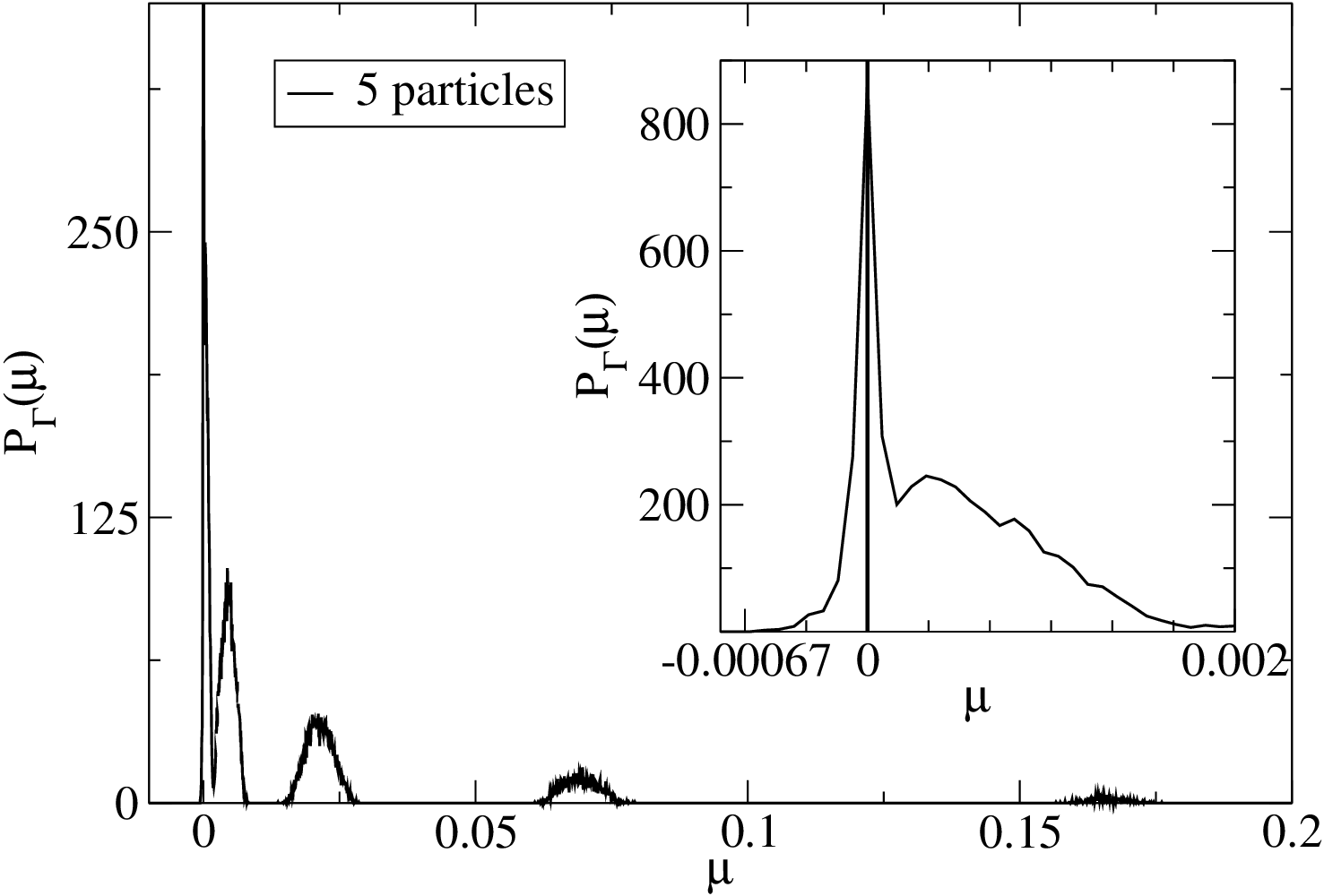}  \\
\includegraphics[width=0.49\linewidth,clip]{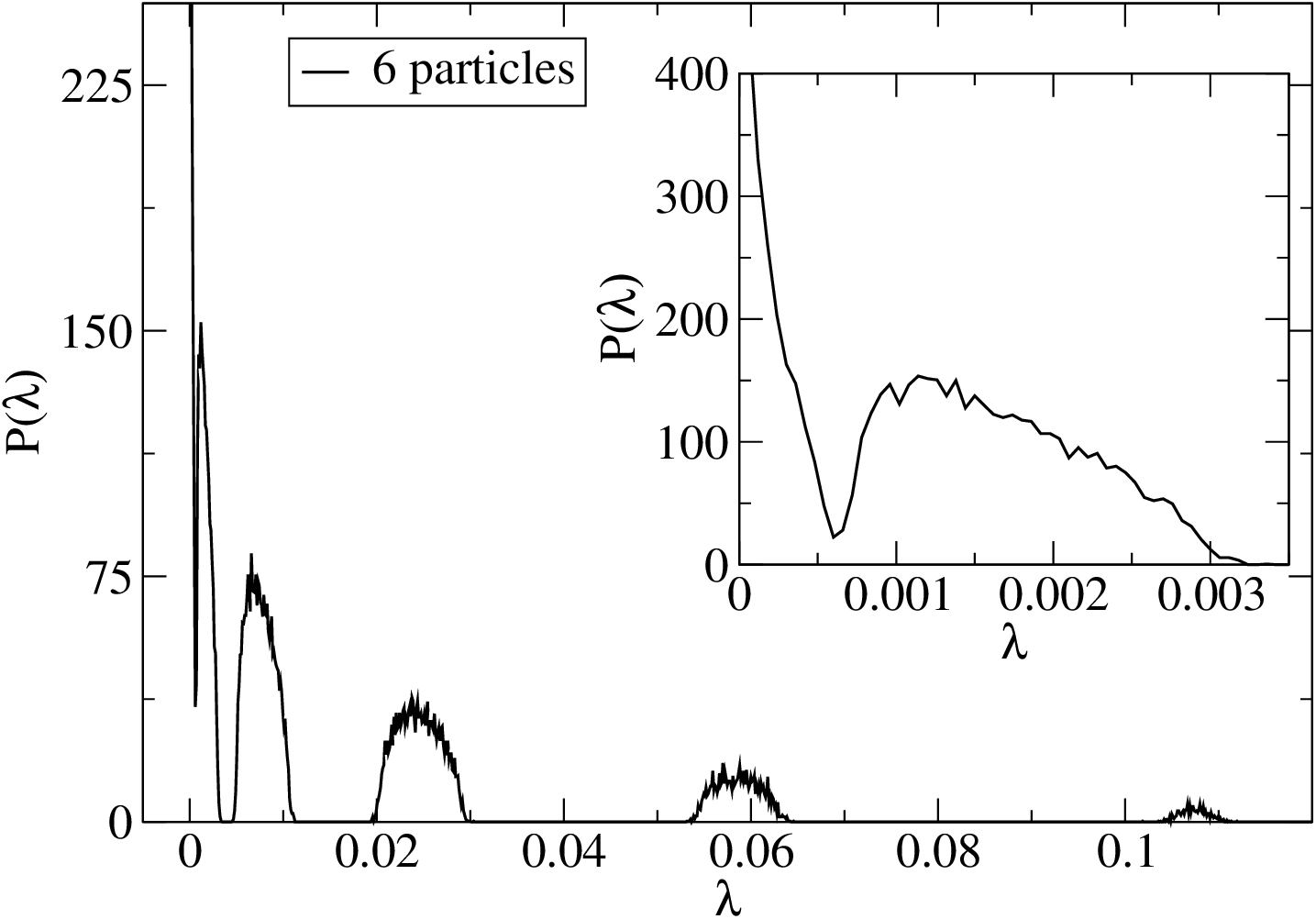}  
\includegraphics[width=0.49\linewidth,clip]{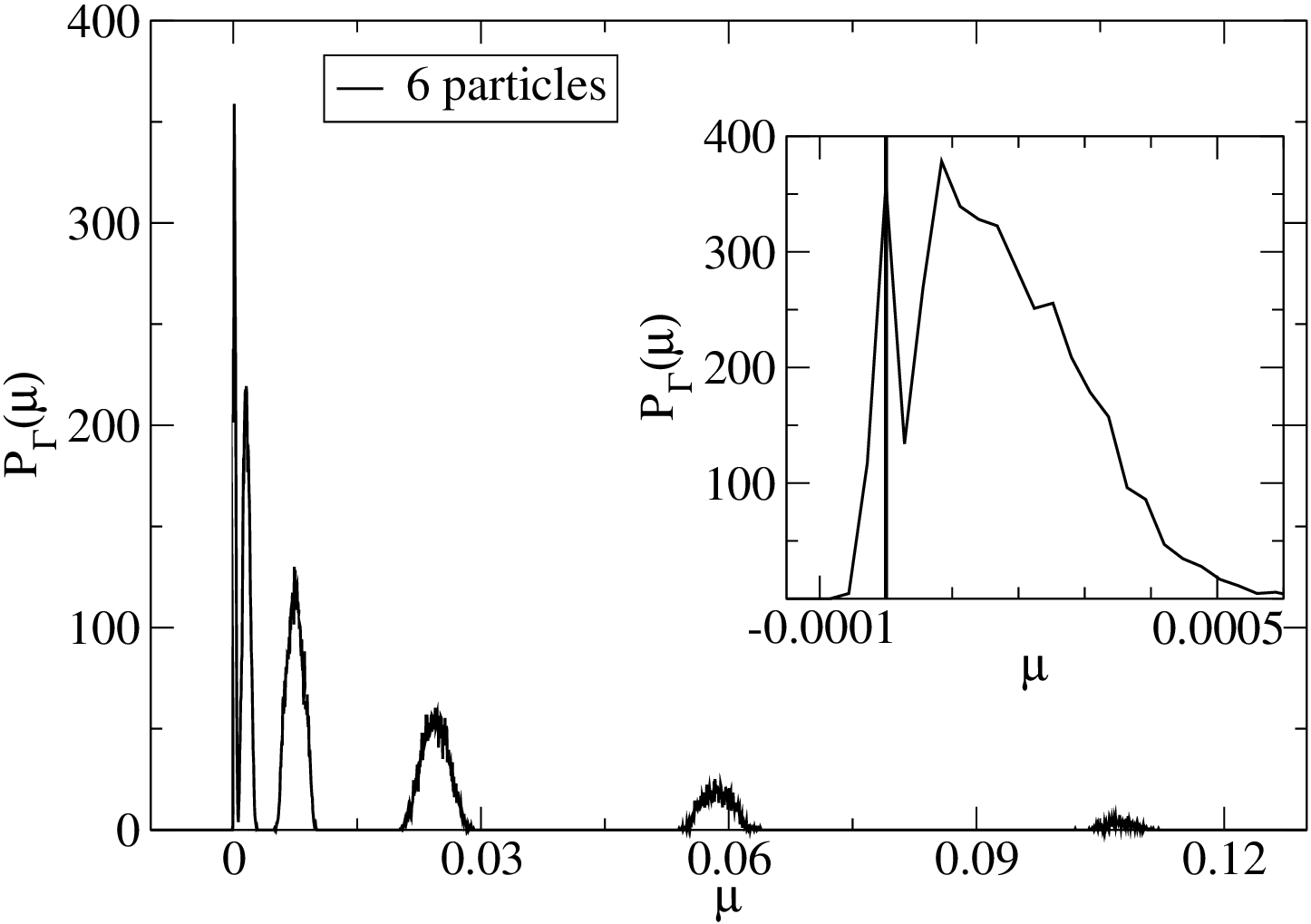}\\
\includegraphics[width=0.49\linewidth,clip]{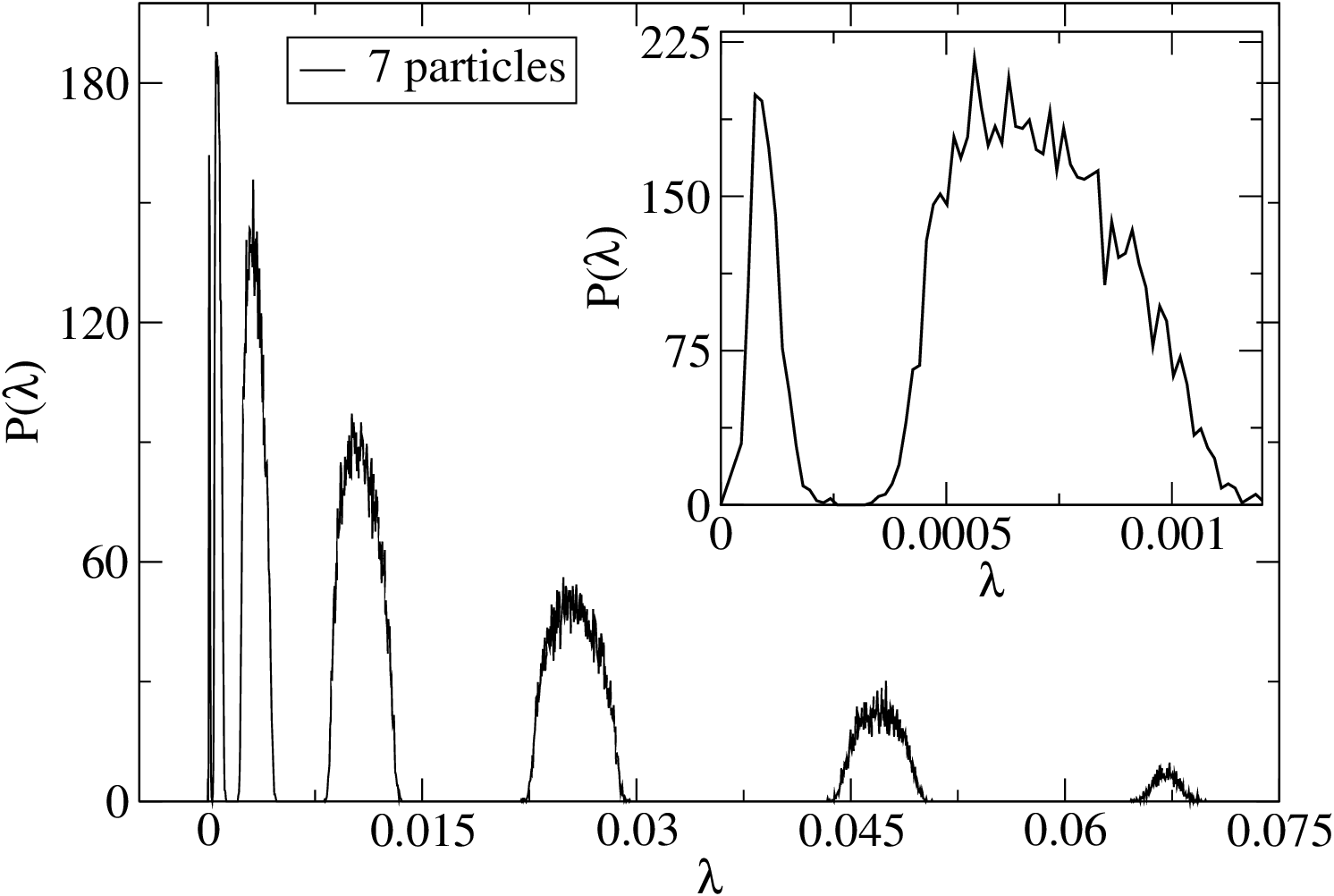}
\includegraphics[width=0.49\linewidth,clip]{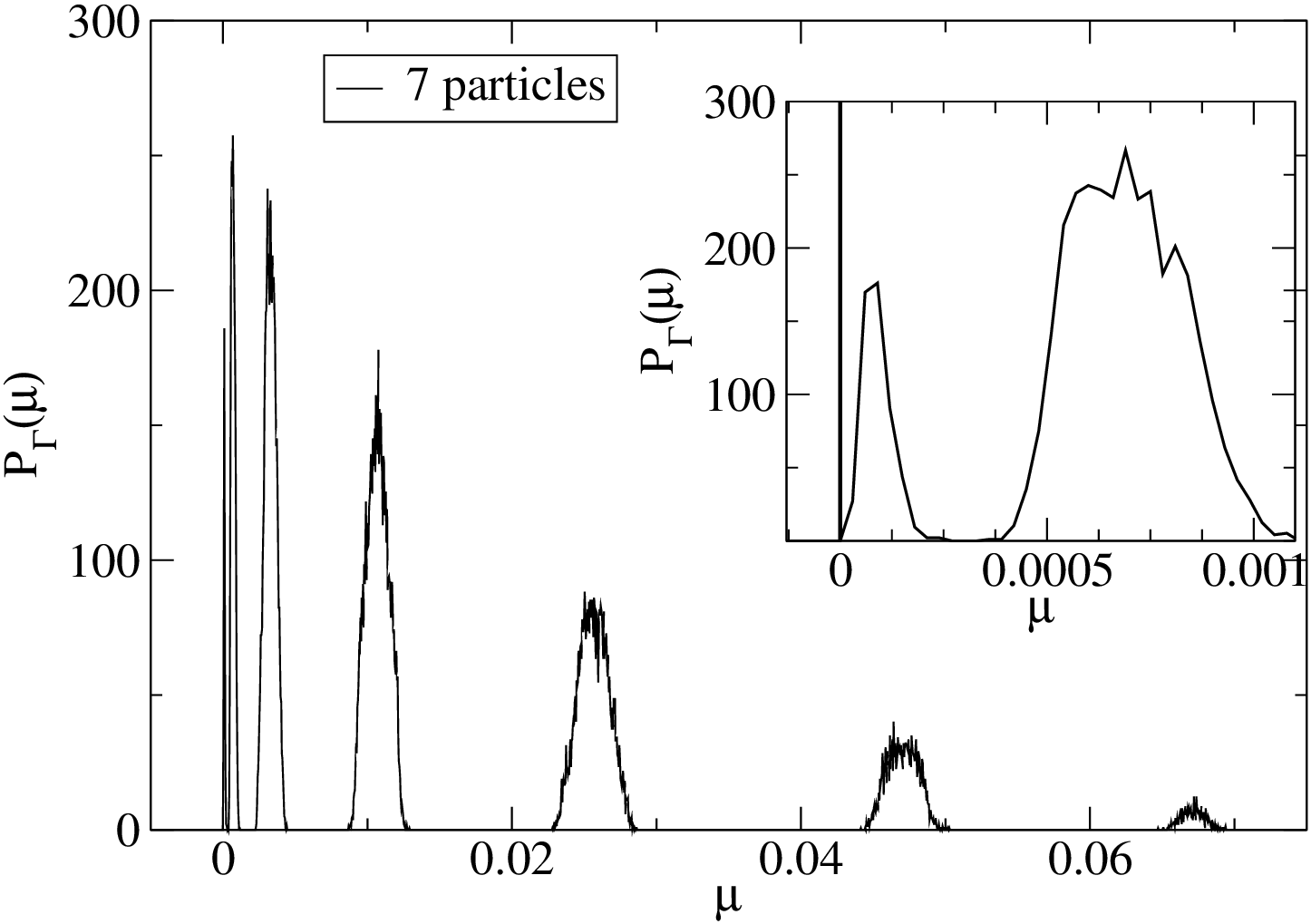}  \\
\caption{Density of states of the reduced density matrix $\rho_A$ (left column) and its partial transpose $\rho_A^{\Gamma}$ (right column). The block length ($m$) is equal to $6$, the total number of qubits ($N$) is $22$ and the particle number ($l$) varies as shown. The insets shows an enlarged view of the  region near the origin of the respective figures. In the insets of the right column a vertical line at the origin has been shown to 
 draw attention to the negative part of the spectrum.}
\label{fig:dos}
\end{figure}

The spectral properties of the reduced density matrix which represents the state of the block whose entanglement is under investigation is of natural interest. Apart from being positive semi-definite
the eigenvalues of the reduced density matrix of a pure random state has a characteristic
distribution or ``density of states", which is discussed further below. In contrast the corresponding
spectrum for the partial transpose need not be positive semi-definite; indeed if the density of states
now of the PT of the reduced density matrix, has support in the negative numbers, the corresponding
state is entangled or NPT. 
The mechanism that is responsible for the transitions pointed to above remains to be fully investigated, however the density of states of the reduced density matrix may be playing a crucial role. 

To elucidate this, as discussed in the Introduction, when $l<m$, that is the number of particles is smaller than the block length, there are many exact zero eigenvalues in the 
density matrix $\rho_A$, in contrast when $l \ge m$ the density matrix becomes of full rank. In fact when $l=m$ a detailed
study of the density matrix shows a density of states that still diverges at 0, while for $l>m$ the density of states vanishes at zero. Negative eigenvalues develop in the  partial transpose of the rank-deficient matrices corresponding to the case $l<m$. When the density of states of $\rho_A$ is bounded away from zero, as in the case of $l>m$, the partial transpose is also bounded away from zero and has typically only positive eigenvalues. In the marginal case when $l=m$ the divergent density of states of $\rho_A$ seems to lead to negative partial transpose. Thus whenever
a density matrix has exact zero eigenvalues, or has a divergent density of states at zero, it will be typically NPT, and hence
entangled. As the number of particles is increased beyond the block size, the spectrum of $\rho_A$ gets bounded away from 
zero and it becomes PPT. Thus the transition seems to originate in the transition of the density of states of the
reduced density matrix, which in turn is due to the rank of the density matrix becoming full at the point of transition.
 However we emphasize that the observations made here are partially numerical and further work on the partial transpose of rank-deficient matrices is necessary to justify them rigorously.

If $|\psi \rangle$ is a full random state of $N$ qubits, mixing all particle numbers together, and let 
the subset $A$ have Hilbert space dimension $d_A$ and the complementary set,  dimension $d_B$ ($d_A \le d_B$).
Then the density of states of the reduced density matrix
of a subset $A$ of the qubits $\rho_A$, if $d_A,d_B \gg1$, will typically be distributed according to the Marcenko-Pastur rule \cite{Marcenko67}:
\begin{equation}
\begin{split}
  f(\lambda)& = \frac{d_A Q}{2\pi} \frac{\sqrt{(\lambda_{max} -\lambda )(\lambda- \lambda_{min})}}{\lambda}\\
{\lambda_{min}^{max}} &= \frac{1}{d_A}\bigg(1+\frac{1}{Q} \pm \frac{2}{\sqrt{Q}}\bigg); \;\; Q=d_B/d_A.
\end{split}
\label{MPfunct}
\end{equation}
In the ``symmetric" case of $Q=1$, the density of states diverges at the origin, else it is bounded away from 
zero. In fact much is known about the distribution of the smallest eigenvalue in the symmetric case, including 
its distribution and average ($1/d_A^3$) \cite{Majumdar09}. 

Corresponding questions for definite particle subspaces are of natural interest, and we present some results here
for the density of states, but only in so far as they pertain to the problem of entanglement transition studied above. Thus in addition 
to the density of states, $P(\lambda)$ of the density matrix $\rho_A$ the density of states, $P_{\Gamma}(\mu)$, of the partial transpose,   $\rho_A^{\Gamma}$ is of interest. In Fig.~(\ref{fig:dos}) is shown the density of states before and after the partial transposition for a case
when there are $N=22$ qubits in all. The block $A$ consists of $m=6$ qubits and the density of states is shown as the particle
number is changed across this value. Looking at the density of states $P(\lambda)$ for the case $l=6$ particles one can see the divergence
at the origin as well as several clumps of eigenvalues. The origin of the clumps is quite easily understood as arising from the individual 
$G_k$ blocks acting as practically independent density matrices, the trace normalization condition being the only constraint amongst them.
These individual blocks then tend to have density of states that are of the nature of the Marcenko-Pastur distribution with suitable 
dimensions. Thus roughly, especially for large $N$, the density of states is pretty much a superposition of such distributions. 

 The eigenvalues $\lambda$ of the extreme nonzero blocks $G_0$ and $G_r$ where $r=m$ or $l$ depending on whether $l\ge m$ or $<m$, are special in the sense that there is only a lone nonzero eigenvalue and they do not follow the Marcenko-Pastur distribution. In any case the ``block" $G_0$ is just a number, while $G_r$ is a matrix for $l<m$ and a number for $l\ge m$.
 For example in the case when $m=2$, these two 
 numbers are the values of $a_{00}$ and $a_{33}$ of Eq.~(\ref{rhoA}), which are seen to be the sum of squares of the normally distributed coefficients. 
 Therefore it is easy to see that in general they are  chi-squared distributed with number of degrees of freedom $d$:  
\begin{eqnarray}
 \frac{1}{2^{d/2}\Gamma(d/2)}\;x^{d/2-1}  e^{-x/2},\;\; x \in [0,\infty)\; \mbox{and}\; d \geq 1,
\label{chi2}
\end{eqnarray} 
where $x= \lambda \mathcal{N}$.
The number of degrees of freedom $d$ depends on whether $l \le m$ or $l>m$. 
In either case for the eigenvalue of $G_0$, $d$ is equal to ${N-m \choose l}$. In the case of the eigenvalue
of the block $G_r$, $d= {N-m \choose l-m}$ when $l \ge m$, and $d={m \choose l}$ for $l < m$. Thus when $l=m$, the number of particles is the
block size, $d=1$ for the eigenvalue of the block $G_r$. Note that when $d=1$ the chi-squared distribution diverges at the origin, unlike the case $d>1$.  Thus although at the transition point $l=m$, the density matrix $\rho_A$ is of full rank, it has a divergent density
of states arising from this eigenvalue. Note that when $l>m$ these lone eigenvalues can never lead to a  divergent density of states. Also from the inequality in Eq.~(\ref{mchoosek}) it follows that there will never be a symmetric
case of the Marcenko-Pastur distribution when $l>m$. Thus indeed this completes the proof that the density of states does not diverge at zero when $l>m$, while it does for $l\le m$.

Not much is known of the density of states of the partial transpose even for the case of full random states, except for a recent mathematical
study \cite{PTrandom} and an ongoing work \cite{BhosaleTomsovicLak} which shows for instance that when a density matrix has a symmetric Marcenko-Pastur distribution ($Q=1$), its partial transpose has a semi-circle distribution. Indeed the density of states of the partial transposed matrix $\rho_A^{\Gamma}$ as shown in Fig.~(\ref{fig:dos})
is not very different from that of the density matrix itself. The important exceptions are cases where the density of states diverges at the origin 
(in the case of $l=5$ and $6$) and the density of states of $\rho_A^{\Gamma}$ clearly has support in the negative numbers, indicating the NPT nature of $\rho_A$. In contrast when there are  $7$ particles  the PT has almost no negative eigenvalues. A much more detailed study of the tails
of these distributions show a very small fraction of negative eigenvalues, indicating that entanglement when present is very rare. This is reflected in the
exponentially small probability of entangled states after the transition ($l > m$). Thus the origin of the entanglement transitions seems to lie in the change of character of the density of states of the
reduced density matrix around zero.

\section{Discussions and conclusion}
While we have focused on the study entanglement {\em within} a block
of qubits, one may also ask if transitions are seen in the
entanglement of the block with the rest of the qubits, say measured by
the von Neumann entropy. Preliminary work not presented here, as well
as from the discussions above we believe that no such transition is
seen. This is due to the fact that von Neumann entropy is fairly
insensitive to the nature of the density of states around the zero
eigenvalue. For instance even in the case of full random states, the
symmetric states (equal bipartitions) do not possess qualitatively different entanglement
entropy from the non-symmetric ones \cite{Page93}.  On the other hand,
entanglement within the block, as measured by the log-negativity (or
the concurrence in the case of two qubits) is sensitive to the
presence of a large number of zero or near zero eigenvalues.

In summary this paper has given definitive evidence of a transition in
entanglement between two qubits as the number of particles is
increased to three. Using log-negativity it is shown that the
following generalization would hold: the entanglement content in $m$
qubits decays algebraically with $N$, the number of qubits, if the
number of particles $l\le m$, and exponentially if $l>m$. Various
exponents in the case of algebraic decay have been analytically
derived for the case of concurrence as well as the log-negativity.
The observation of a transition is further strengthened by studying
the density of states of the reduced density matrix and its partial
transpose. It is shown that the rank of the density matrix is not full
till the number of particles is precisely equal to the block size; and
that even at exactly the marginal case, the density of states of the
reduced density matrix diverges at zero, although it is of full
rank. The exact zero eigenvalues or the large number of very small
ones seem to translate on partial transpose to negative eigenvalues,
thus leading to typically entangled states. This is the case as long
as the number of particles is less than or equal to the block
size. Otherwise the density of states vanishes at zero and leads to a
predominantly positive partial transpose, which results in the
exponentially small entanglement.

The question of whether the transition studied is also observed on
using other entanglement measures is a natural and interesting one. 
It is quite easy to see that the negativity measure (rather than 
the log-negativity studied here) also undergoes such a transition. Note that unlike the 
log-negativity measure which is not convex but is nevertheless an entanglement monotone \cite{Plenio05}, the 
negativity measure is both convex and an entanglement monotone \cite{vidal02}. 
While many other measures, such as distillable entanglement \cite{nielsenbook}, are difficult to compute,
it seems very plausible that the transition is indeed independent of the particular
measure used. This is strengthened by the study of the spectra of the reduced
density matrix and its partial transpose, and the fact that the transition may well
have its origins in the behavior of their density of states.

\begin{acknowledgments}
It is a pleasure to thank Steven Tomsovic for many useful discussions. This work was partially
supported by funding from the DST, India, under the project SR/S2/HEP-012/2009.

\end{acknowledgments}

\bibliography{refs}

\end{document}